\documentclass[runningheads]{llncs}
\usepackage{graphicx}
\usepackage{multirow}
\usepackage{caption}
\usepackage{subcaption}
\usepackage{xcolor}
\usepackage{array,booktabs,ragged2e}
\usepackage{mathtools}
\newcommand{\comment}[1]{}
\newcolumntype{R}[1]{>{\RaggedLeft\arraybackslash}p{#1}}
\begin{document}
\title{3D Semantic Segmentation of  Brain Tumor for Overall Survival Prediction}
\titlerunning{3D Semantic Segmentation of  Brain Tumor for Overall Survival Prediction}
\author{Rupal R. Agravat\inst{1}\orcidID{0000-0003-1995-4149} \and
Mehul S. Raval\inst{1}\orcidID{0000-0002-3895-1448}}
\authorrunning{Agravat R. and Raval M.}
\institute{Ahmedabad University, Ahmedabad, Gujarat, India \\
\email{rupal.agravat@iet.ahduni.edu.in}\\
\email{mehul.raval@ahduni.edu.in}}
\maketitle              % typeset the header of the contribution
\begin{abstract}
Glioma, a malignant brain tumor, requires immediate treatment to improve the survival of patients. The heterogeneous nature of Glioma makes the segmentation difficult, especially for sub-regions like necrosis, enhancing tumor, non-enhancing tumor, and edema. Deep neural networks like full convolution neural networks and an ensemble of fully convolution neural networks are successful for Glioma segmentation. The paper demonstrates the use of a 3D fully convolution neural network with a three-layer encoder-decoder approach. The dense connections within the layer help in diversified feature learning. The network takes 3D patches from $T_{1}$, $T_{2}$, $T_{1}c$, and FLAIR modalities as input. The loss function combines dice loss and focal loss functions. The Dice similarity coefficient for training and validation set is 0.88, 0.83, 0.78 and 0.87, 0.75, 0.76 for the whole tumor, tumor core and enhancing tumor, respectively. The network achieves comparable performance with other state-of-the-art ensemble approaches. The random forest regressor trains on the shape, volumetric, and age features extracted from ground truth for overall survival prediction. The regressor achieves an accuracy of 56.8\% and 51.7\% on the training and validation sets. 

\end{abstract}

\section{Introduction}

Early-stage brain tumor diagnosis can lead to proper treatment planning, which improves patient survival chances. Out of all types of brain tumors, Glioma is one of the most life-threatening brain tumors. It occurs in the glial cells of the brain. Depending on its severity and aggressiveness, Glioma has grades ranging from I to IV. Grade I, II are Low-Grade Glioma (LGG), and grade III and IV are High-Grade Glioma (HGG). It can further be divided into constituent structures like - necrosis, enhancing tumor, non-enhancing tumor, and edema. The core consists of necrosis, enhancing tumor, non-enhancing tumor. In most cases LGG does not contain enhancing tumor, whereas HGG contains necrosis, enhancing, and non-enhancing structures. Edema occurs from infiltrating tumor cells and biological response to the angiogenic and vascular permeability factors released by the spatially adjacent tumor cells\cite{akbari2014pattern}. Non-invasive Medical Resonance Imaging (MRI) is the most advisable imaging technique as it captures the functioning of soft tissue adequately compared to other imaging techniques. MR images are prone to inhomogeneity introduced by the surrounding magnetic field, which introduces the artifacts in the captured image. Besides, the appearance of various brain tissues is different in various modalities. Such issues increase the time in the study of the image.

The treatment planning is highly dependent on the accurate tumor structure segmentation, but due to heterogeneous nature of Glioma, the segmentation task becomes difficult. Furthermore, the human interpretation of the image is non-reproducible as well as dependent on expertise. It requires computer-aided MR image interpretation to locate the tumor. Also, even if the initially detected tumor is completely resected, such patients have poor survival prognosis, as metastases may still redevelop. It leads to an open question about overall survival prediction.

Authors in \cite{agravat2018deep} discussed the basic, generative, and discriminative techniques for brain tumor segmentation. Nowadays, Deep Neural Network (DNN) has gained more attention for the segmentation of biological images. The Convolution Neural Networks (CNN), like DeepMedic \cite{kamnitsas2017efficient}, U-net \cite{ronneberger2015u}, V-Net \cite{milletari2016v}, SegNet \cite{badrinarayanan2017segnet}, ResNet \cite{he2016deep}, DenseNet \cite{iandola2014densenet} give state-of-the-art results for semantic segmentation. Out of all these methods, U-net is a widely accepted end-to-end segmentation architecture for brain tumors. U-net is an encoder-decoder architecture, which reduces the size of feature maps to half and doubles the number of feature maps at every encoder layer. The process is reversed at every decoder layer. The skip connections between the peer layers of U-net help in proper feature reconstruction.

\subsection{Literature Review: BraTS 2019}

\subsubsection{Segmentation}

Authors in \cite{jiang2020two} used the ensemble of twelve encoder-decoder models, where each model is made up of a cascaded network. The first network in a model finds the coarse segmentation, which was given as input in the second network, and the input images to predict all the labels. The network losses combine at different stages for better network parameter tuning. The Dice Similarity Coefficient (DSC) for the validation set is 0.91, 0.87, 0.80 for Whole Tumor (WT), Tumor Core (TC), and Enhancing Tumor (ET), respectively.

In \cite{zhao2019bag}, authors applied various data processing methods, network design methods, and optimization methods to learn the segmentation labels at every iteration. The student models combined at the teacher-level model with successive output merging. The loss function is the combination of dice loss and cross-entropy loss for the networks trained on various input patch sizes. The method achieved the DSC of 0.91, 0.84, and 0.75 for WT, TC, and ET, respectively.

The approach demonstrated in \cite{mckinley2019triplanar} used thirty Heteroscedastic classification models to find the variance of all the models for the ensemble. The focal loss forms the loss function. Various post-processing techniques were applied to fine-tune the network segmentation. The DSC achieved for the approach was 0.91, 0.83, and 0.77 for WT, TC, and ET.

Authors in \cite{starke2019integrative} used fifteen 2D FCNN models working in parallel on axial, coronal and sagittal views of images. The approach used in \cite{guo2019brain} focused on the attention mechanism applied for the ensemble of four 3D FCNN models. In \cite{kotowski2019detection}, authors used the ensemble of four 2D FCNN models working on different sets of images based on the size of the tumor. 

\subsubsection{Survival Prediction}

Authors in \cite{agravat2019brain} implemented 2D U-net with dense modules at the encoder part and convolution modules at the decoder part along with focal loss function at training time. The segmentation results fed into Random Forest Regressor(RFR) to predict the Overall Survival (OS) of the patients. The RFR trains on the age, shape, and volumetric features extracted from the ground truth provided with the training dataset. They achieved 58.6\% OS accuracy on the validation set.

Authors in \cite{wang2019automatic} used vanilla U-net and U-net with attention blocks to make the ensemble of six models based on various input patches and the presence/absence of attention blocks. The linear regressor trains selected radiomics features along with the relative invasiveness coefficient. The DSC achieved on the validation set was 0.90, 0.83, and 0.79 for WT, TC, and ET, respectively, and the OS accuracy was 59\%.

Authors in \cite{feng2019brain} implemented the ensemble of six models, which are the variation of U-net with different patch sizes, feature maps with several layers in the encoder-decoder architecture. For OS prediction, six features were extracted from the segmentation results to train the linear regression. The DSC achieved on the validation set was 0.91, 0.80, and 0.74 for WT, TC, and ET, and the OS accuracy was 31\%. 

Authors in \cite{wang20193d} used the U-net variation, where the additional branch of prediction uses Variational Encoder. The OS prediction used the volumetric and age features to train ANN with two layers, each with 64 neurons. 

Except the approach demonstrated in \cite{agravat2019brain,wang20193d}, all the other approaches use an ensemble of the segmentation prediction networks. There are certain disadvantages of ensemble approaches: (1) ensemble methods are usually computationally expensive. Therefore, they add learning time, and memory constraints to the problem, (2) using ensemble methods reduces the model interpretability due to increased complexity and makes it very difficult to understand.

The focus of this paper is to develop a robust 3D fully convolutional neural network (FCNN) for tumor segmentation along with RFR \cite{agravat2019brain} to predict OS of high grade glioma (HGG) patients. The remaining paper is as follows: section \ref{dataset} focuses on the BraTS 2020 dataset, section \ref{Proposed Method} demonstrates the proposed methods for tumor segmentation and OS prediction, section \ref{Implementation Details} provides implementation details, and section \ref{Results} discusses the results followed by the conclusion and future work.

\section{Dataset}
\label{dataset}

The dataset \cite{bakas2017advancing,bakas2018identifying,menze2014multimodal} contains 293 HGG and 76 LGG pre-operative scans in four Magnetic Resonance Images (MRI) modalities, which are $T_{1}$, $T_{2}$, $T_{1}c$ and FLAIR. One to four raters have segmented the images using the same annotation tool to prepare the ground truths. The annotations were approved by experienced neuro-radiologists \cite{bakas2017segmentation,bakas2017segmentation1}. Annotations have the enhancing tumor (ET label 4), the peritumoral edema (ED label 2), and the necrotic and non-enhancing tumor core (NCR/NET label 1). The $T_{2}$, $T_{1}c$ and FlAIR images of a single scan are co-registered with the anatomical template of $T_{1}$ image of the same scan. All the images are interpolated to the same resolution (1mm x 1mm x 1mm), and skull-stripped. Features like age, survival days and resection status for 237 HGG scans are provided separately for OS. The validation and test datasets consist of 125 and 166 MRI scans respectively, with the same pre-processing. The dataset includes age and resection status for all the sets along with survival days for the training set.

\section{Proposed Method}
\label{Proposed Method}

\subsection{Task 1: Tumor Segmentation}
\begin{figure}
	\centering
\includegraphics[width=\textwidth,height=7cm]{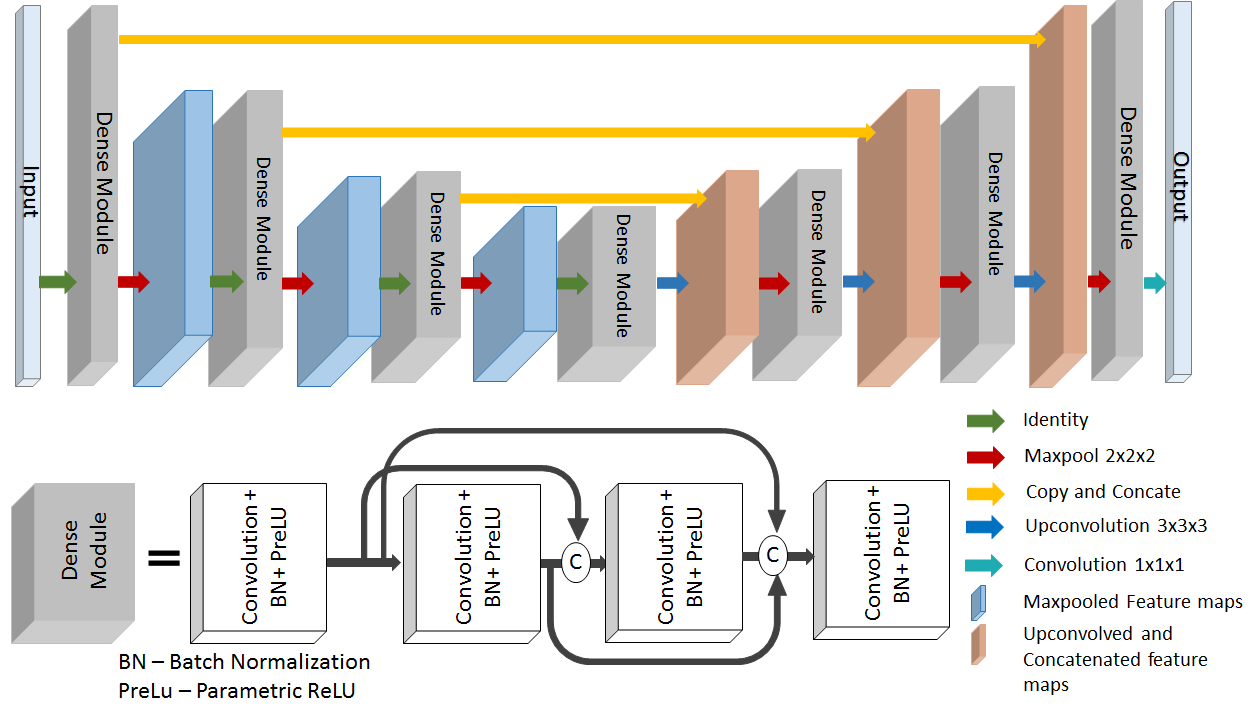}
\caption{The proposed 3D encoder-decoder FCNN.} \label{fig1}
\end{figure}

A FCNN provides end-to-end semantic segmentation for the input of the arbitrary size and learns global information related to it. Moreover, the 3D FCNN gathers spatial relationships between the voxels. Our network is an extension of our previous work proposed in \cite{agravat2019brain}, where the network had poorly performed on validation set and test set. The network overfitted the training set and could not learn the 3D voxel relationship. The proposed work includes 3D modules with increased network depth. The network uses three-layer encoder-decoder architecture with dense connections between the convolution layers and skip-connections across peer layers. The network is as shown in Fig. \ref{fig1}. The Batch Normalization (BN) and Parametric ReLU (PReLU) activation function follow the convolution layer in each dense module.  

Dense connections between the layers in the dense module allows to obtain additional inputs(collective knowledge) from all earlier layers and passes on its feature-maps to all subsequent layers. It allows the gradient to flow to the earlier layers directly, which provides in-depth supervision on preceding layers by the classification layer. Also, dense connections provide diversified features to the layers, which leads to detailed pattern identification capabilities. The layers generate 64, 128 and 256 feature maps and the bottleneck layer generates 512 feature maps. The 1x1x1 convolution at the end generates a single probability map for multi-class classifications.

Brain tumor segmentation task deals with a highly imbalanced dataset where tumorous slices are less than non-tumorous slices; such an imbalance dataset reduces network accuracy. Two approaches deal with the issue: 1) The patch-based input to the network guarantees that the network does not overlearn the background voxels, 2) the network trains with the combination of the following loss functions.

\begin{itemize}
\item Soft Dice Loss: is a measure to find overlap between two regions. 
\[
Soft\:Dice\:Loss = 1 - \frac{2\: \sum_{voxels}{y_{true}}{y_{pred}}}{\sum_{voxels}{y_{pred}}^{2}+\sum_{voxels}{y_{true}}^{2}}
\]
${y_{true}}$ represents ground truth and ${y_{pred}}$ represents network output probability. The dice loss function directly considers the predicted probabilities without converting them into binary output. The numerator provides standard correct predictions between input and target, whereas the denominator provides individual separate correct predictions. This ratio normalizes the loss according to the target mask and allows learning even from the minimal spatial representation of the target mask.

\item Focal Loss\cite{lin2017focal}: It is dependent on the network probability ${p_{t}}$. It balances negative and positive samples by tuning weighting parameter $\alpha$. It also deals with easy and hard examples by tuning the focusing parameter $\gamma$.

\begin{equation}
FL({p_{t}}) = -{\alpha_{t}}\:(1 - {p_{t}})^{\gamma}\:log({p_{t}})
\end{equation}

The modulating factor $(1 - {p_{t}})^{\gamma}$ adjusts the rate at which easy examples are down-weighted.
\end{itemize}

\subsection{Task 2: Overall Survival Prediction}
OS prediction deals with predicting the number of days for which patients survive after the tumor is resected and proper post-operative treatment given.  We have used the following features to train RFR:
\begin{itemize}
\item \textbf{Statistical Features}: the amount of edema, amount of necrosis, amount of enhancing tumor, the extent of tumor and proportion of tumor.
\item \textbf{Radiomic Features\cite{van2017computational} for necrosis}: Elongation, flatness, minor axis length, primary axis length, 2D diameter row, 2D diameter column, sphericity, surface area, 2D diameter slice, 3D diameter.
\item \textbf{Age}(available with BraTS dataset).
\end{itemize}

Necrosis plays a significant role in the treatment of tumors. Gross Total Resection(GTR) of necrosis is comparatively easy vis a vis enhancing tumor. Considering this, shape features of necrosis are extracted using a radiomics package \cite{van2017computational}. In addition to these features, whole tumor statistical features from the segmentation results and age are considered to train RFR.

\section{Implementation Details}
\label{Implementation Details}

\subsection{Pre-processing}

Pre-processing boosts network training and improves performance. All four modality images are biased field corrected followed by denoising, and Z-score normalization on individual MR sequence is applied where each sequence was subtracted by its mean from the data and divided by its standard deviation. Data augmentation happens by flipping the patches around the vertical axis.

\subsection{Training}

Input to the network is patches of size 64x64x64 from four modalities (T1, T2, T1c, FLAIR). The network trains on the entire training image dataset. The network uses the combination of two loss functions: 1) dice loss function and 2) focal loss function with $ \alpha = 1 $ and $ \gamma = 2 $. The network trains for 610 epochs with batch size 1. The batch normalization with batch size 1 does not speed up the network learning but helps during testing, where it uses the statistics of the batch rather than the running statistics. This layer normalization approach \cite{ba2016layer} normalizes each feature map with its own mean and standard deviation. 

The sliding window approach provides the output for each subject. The stride size is reduced to half of the training window size to overcome the boundary voxels' unstable prediction issue. The output of the original patch and flipped patch is predicted and averaged to generate the final output. The prediction of a single image takes around one minute. 

\subsection{Post-processing}
The post-processing includes two methods: 1) The connected component analysis (CCA) removes the tumor with a volume less than thousand voxels, 2) enhancing tumor is formed in the surrounding of the necrosis. and its size cannot be very small in HGG. Such small size enhancing tumor is converted to necrosis. The empirically chosen threshold for the conversion is three hundred.
 
\section{Results}
\label{Results}

\subsection{Segmentation}

\comment{
%The achieved DSC, Hausdorff95, Sensitivity and Specificity for %training set are in Table \ref{tab1} and Table \ref{tab2}, for the %validation set in Table \ref{tab3} and Table \ref{tab4}, and for test set are in Table \ref{tab5} and Table \ref{tab6}. The DSC achieved on training, validation and test sets shows that the network is not overfitting and gives consistent performance on all the datasets.
}

The achieved DSC, sensitivity, specificity and Hausdorff95 for training set is in Table \ref{tab1}, and for the validation set in Table \ref{tab2}. The results shows the improvement with post-processing, hence the test set results are generated with post-processing as shown in Table \ref{tab3}.

\begin{table}[h]
	\centering
	\caption{Various evaluation measures for BraTS 2020 training set.}
	\label{tab1}
	 \begin{tabular}{cp{2cm}p{1.2cm}p{1.2cm}p{1.2cm}p{1.2cm}p{1.2cm}p{1.2cm}}
		\toprule
		\multirow{2}{*}{Evaluation Measure} &
		\multicolumn{1}{c}{\multirow{2}{*}{Statistics}} &
		\multicolumn{3}{c}{Without post-process} &
		\multicolumn{3}{c}{With post-process} \\
		\midrule
		&
		\multicolumn{1}{c}{} &
		\multicolumn{1}{c}{ET} &
		\multicolumn{1}{c}{WT} &
		\multicolumn{1}{c}{TC} &
		\multicolumn{1}{c}{ET} &
		\multicolumn{1}{c}{WT} &
		\multicolumn{1}{c}{TC} \\
		\midrule
		\multirow{5}{*}{DSC} & Mean & 0.757 & 0.881 & 0.831 &  0.782 & 0.882 & 0.832 \\
		& StdDev     &  0.267 &	0.115 &	0.191 &  0.246 & 	0.116 &	0.191 \\
		& Median     &  0.863 &	0.919 &	0.908 &  0.872 &	0.919 &	0.910 \\
		& 25quantile &  0.751 &	0.860 &	0.802 &  0.779 & 0.862 & 0.805  \\
		& 75quantile &  0.906 &	0.941 &	0.943  &  0.912 & 0.943 & 0.944  \\
		\midrule
		\multirow{5}{*}{Sensitivity} & Mean       &  0.759 & 0.846 &	0.802  &  0.782 & 0.844 &	0.801   \\
		& StdDev     &  0.272 &	0.156 &	0.209  &  0.252 &	0.158 &	0.209  \\
		& Median     &  0.858 &	0.896 &	0.883   & 0.864 &	0.896 &	0.883 \\
		& 25quantile &  0.748 &	0.797 &	0.763  &  0.762 &	0.795 &	0.763 \\
		& 75quantile &  0.921 &	0.946 &	0.931  &  0.925 &	0.945 &	0.931 \\
		\midrule
		\multirow{5}{*}{Specificity} & Mean       & 0.999 &	0.999 &	0.999  &  0.999 &	0.999 &	0.999  \\
		& StdDev     & 0.000 & 	0.000 &	0.000  & 0.000 & 	0.000 &	0.000  \\
		& Median     &  0.999 &	0.999 &	0.999  &  0.999 &	0.999 &	0.999  \\
		& 25quantile &  0.999 &	0.999 &	0.999  &  0.999 &	0.999 &	0.999  \\
		& 75quantile &  0.999 &	0.999 &	0.999  &  0.999 &	0.999 &	0.999 \\
		\midrule
		\multirow{5}{*}{Hausdorff95} & Mean &  31.531	 & 06.508 &	07.275  &  29.274 & 06.232 & 06.999  \\
		& StdDev     &  95.090 & 08.588 & 10.946 &  94.956 & 	08.134 &	20.507  \\
		& Median     &  01.732 &	03.606 &	03.606  &  01.414 &	03.464 &	03.317  \\
		& 25quantile & 01.414 & 02.236 &	02.000 &  01.000 &	02.236 &	02.000  \\
		& 75quantile &  04.123 & 07.071 & 08.602 &  03.162 & 06.939 &	07.874 \\
		\bottomrule
	\end{tabular}
\end{table}

\begin{table}[h]
	\centering
	\caption{Various evaluation measures for BraTS 2020 validation set.}
	\label{tab2}
	\begin{tabular}{cp{2cm}p{1.2cm}p{1.2cm}p{1.2cm}p{1.2cm}p{1.2cm}p{1.2cm}}
		\toprule
		\multirow{2}{*}{Evaluation Measure} &
		\multicolumn{1}{c}{\multirow{2}{*}{Statistics}} &
		\multicolumn{3}{c}{Without post-process} &
		\multicolumn{3}{c}{With post-process} \\
		\midrule
		&
		\multicolumn{1}{c}{} &
		\multicolumn{1}{c}{ET} &
		\multicolumn{1}{c}{WT} &
		\multicolumn{1}{c}{TC} &
		\multicolumn{1}{c}{ET} &
		\multicolumn{1}{c}{WT} &
		\multicolumn{1}{c}{TC} \\
		\midrule
		\multirow{5}{*}{DSC} & Mean & 0.686 & 0.876 & 0.725 &  0.763 &	0.873 &	0.753 \\
		& StdDev     &  0.307 & 0.093 &	0.284 &  0.259 & 0.098 & 0.263  \\
		& Median     &  0.835 &	0.914 &	0.866 & 0.852	& 0.908 & 0.878 \\
		& 25quantile &  0.617 &	0.863 &	0.595 &  0.751	& 0.856 & 0.711  \\
		& 75quantile & 0.889 &	0.932 &	0.924  &  0.899 & 0.935 & 0.926 \\
		\midrule
		\multirow{5}{*}{Sensitivity} & Mean &  0.704 & 0.858 &	0.674 & 0.759 &	0.847 &	0.713   \\
		& StdDev     & 0.322 &	0.138 &	0.312  &  0.273 & 0.149 & 0.288  \\
		& Median     &  0.853 &	0.901 &	0.819   & 0.852 & 0.897 & 0.841 \\
		& 25quantile &  0.595 &	0.826 &	0.432  &  0.715 & 0.809 & 0.606 \\
		& 75quantile &  0.926 &	0.952 &	0.915  &  0.933 & 0.954 & 0.921 \\
		\midrule
		
		\multirow{5}{*}{Specificity} & Mean       & 0.999 &	0.999 &	0.999  &  0.999 &	0.999 &	0.999  \\
		& StdDev     & 0.000 & 	0.000 &	0.000  & 0.001 & 	0.000 &	0.000  \\
		& Median     &  0.999 &	0.999 &	0.999  &  0.999 &	0.999 &	0.999  \\
		& 25quantile &  0.999 &	0.999 &	0.999  &  0.998 &	0.999 &	0.999  \\
		& 75quantile &  0.999 &	0.999 &	0.999  &  0.999 &	0.999 &	0.999 \\
		\midrule
			
		\multirow{5}{*}{Hausdorff95} & Mean  &  43.635	& 09.475 &	14.538  &  27.704	& 07.038	& 10.873  \\
		& StdDev     &  109.143 & 15.215 & 38.067 &  90.918 & 09.348	& 33.823  \\
		& Median     &  02.828 &	04.000 & 05.099  &  02.236 & 03.742 & 04.690 \\
		& 25quantile & 01.414 &	02.236 &	02.236 &  01.414 & 02.449 & 02.236  \\
		& 75quantile &  10.770 & 07.550 & 10.724 &  04.242 & 06.480 & 11.045 \\
		\bottomrule
		\end{tabular}
\end{table}

\begin{table}[h]
	\centering
	\caption{Various evaluation measures for BraTS 2020 test set.}
	\label{tab3}
	\begin{tabular}{cp{2cm}p{1.2cm}p{1.2cm}p{1.2cm}}
		\toprule
		\multirow{2}{*}{Evaluation Measure} &
		\multicolumn{1}{c}{\multirow{2}{*}{Statistics}} &
		\multicolumn{3}{c}{With Post-process} \\
		\midrule
		&
		\multicolumn{1}{c}{} &
		\multicolumn{1}{c}{ET} &
		\multicolumn{1}{c}{WT} &
		\multicolumn{1}{c}{TC} \\
		\midrule
		\multirow{5}{*}{DSC} & Mean & 0.779 &	0.875 &	0.815 \\
		& StdDev     &  0.232 & 0.112 & 0.250 \\
		& Median     &  0.847	& 0.910 & 0.913 \\
		& 25quantile &  0.760	& 0.855 & 0.833  \\
		& 75quantile & 0.908 & 0.935 & 0.948   \\
		\midrule
		\multirow{5}{*}{Sensitivity} & Mean &  0.809 &	0.863 &	0.817 \\
		& StdDev     & 0.245 & 0.136 & 0.244 \\
		& Median     &  0.896 & 0.910 & 0.908 \\
		& 25quantile &  0.785 & 0.823 & 0.795 \\
		& 75quantile &  0.940 & 0.947 & 0.952 \\
		\midrule	
		\multirow{5}{*}{Specificity} & Mean       & 0.999 &	0.999 &	0.999  \\
		& StdDev     & 0.000 & 	0.001 &	0.001  \\
		& Median     &  0.999 &	0.999 &	0.999   \\
		& 25quantile &  0.999 &	0.999 &	0.999   \\
		& 75quantile &  0.999 &	0.999 &	0.999   \\
		\midrule
		\multirow{5}{*}{Hausdorff95} & Mean  &  27.078	& 08.302	& 21.611  \\
		& StdDev     &  92.554 & 30.007 & 74.651   \\
		& Median     &  01.732 & 03.464 & 03.162  \\
		& 25quantile & 01.103 & 02.000 & 01.799   \\
		& 75quantile &  02.734 & 06.164 & 07.858  \\
		\bottomrule
	\end{tabular}
\end{table}

\begin{table}[h]
	\centering
	\caption{DSC comparison with state-of-art ensemble approaches.}
	\label{tab9}
	\begin{tabular}{p{1.5cm}p{3cm}p{3cm}p{1.5cm}p{1.5cm}p{1.5cm}}
		\toprule
		\multirow{2}{*}{Ref.} & \multirow{2}{*}{Type of Network} & \multirow{2}{*}{No. of Networks} & \multicolumn{3}{c}{DSC} \\
		&   &   & ET & WT & TC \\
		\midrule
		\cite{guo2019brain} & \multicolumn{1}{c}{3D FCNN} & \multicolumn{1}{c}{4} & 0.67 & 0.87 & 0.73 \\
		\midrule
		\cite{kotowski2019detection} & \multicolumn{1}{c}{2D FCNN} & \multicolumn{1}{c}{4} & 0.68 & 0.84 & 0.73 \\
		\midrule
		\cite{starke2019integrative} & \multicolumn{1}{c}{2D FCNN} & \multicolumn{1}{c}{15} & 0.71 & 0.85 & 0.71 \\
		\midrule
		\textbf{Proposed} & \multicolumn{1}{c}{3D FCNN} & \multicolumn{1}{c}{\textbf{1}} & \textbf{0.76} &	0.87 &	\textbf{0.75} \\	
		\bottomrule
	\end{tabular}
\end{table}

\begin{figure}[h]
\centering
\begin{subfigure}{.21\textwidth}
\includegraphics[width=3cm,height=3cm]{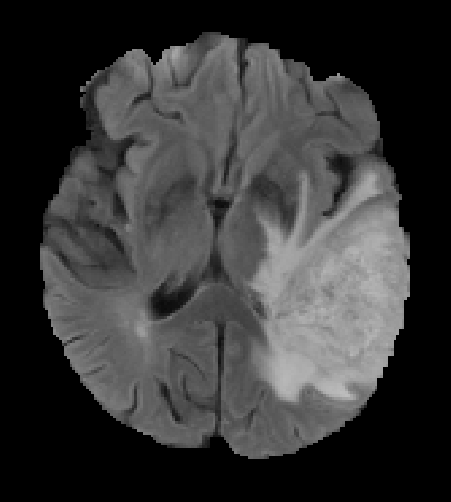}
\caption{}
\end{subfigure} \hfill
\centering
\begin{subfigure}{.21\textwidth}
\includegraphics[width=3cm,height=3cm]{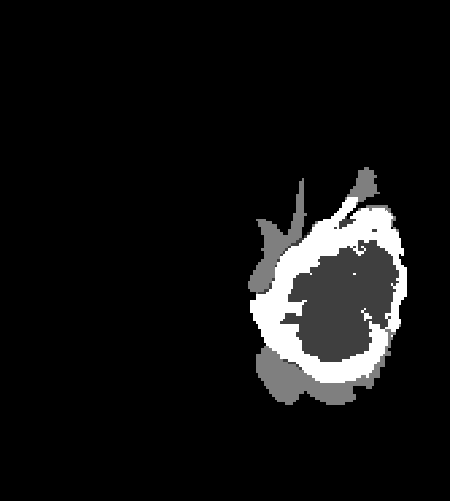}
\caption{}
\end{subfigure} \hfill
\centering
\begin{subfigure}{.21\textwidth}
\includegraphics[width=3cm,height=3cm]{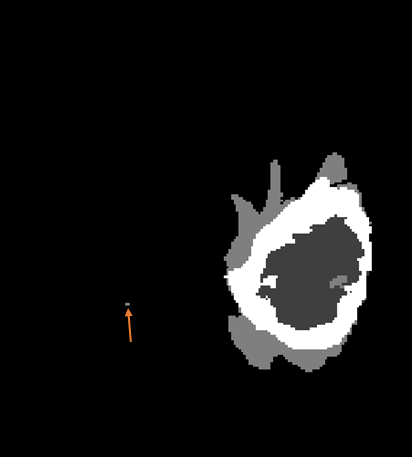}
\caption{}
\end{subfigure} \hfill
\centering
\begin{subfigure}{.21\textwidth}
\includegraphics[width=3cm,height=3cm]{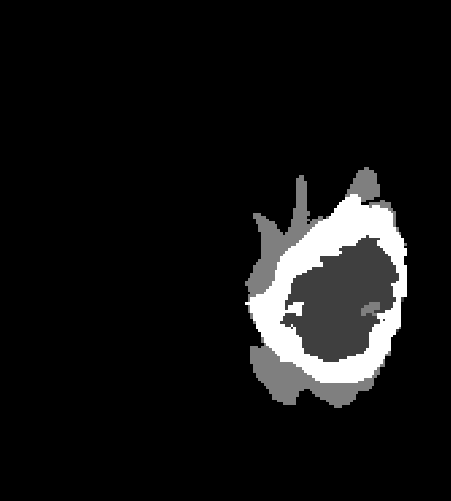}
\caption{}
\end{subfigure}
\caption{Correct segmentation results of the network (a) FLAIR slice (b) Ground truth (c) Segmentation without post-processing (d) Segmentation after post-processing.}
\label{fig2}
\end{figure}

\begin{figure}[h]
\centering
\begin{subfigure}{.21\textwidth}
\includegraphics[width=3cm,height=3cm]{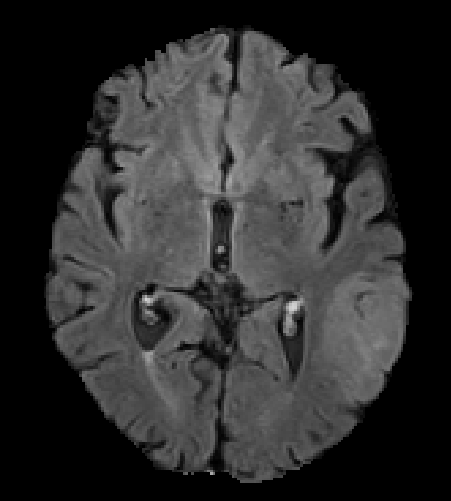}
\caption{}
\end{subfigure}  \hfill
\centering
\begin{subfigure}{.21\textwidth}
\includegraphics[width=3cm,height=3cm]{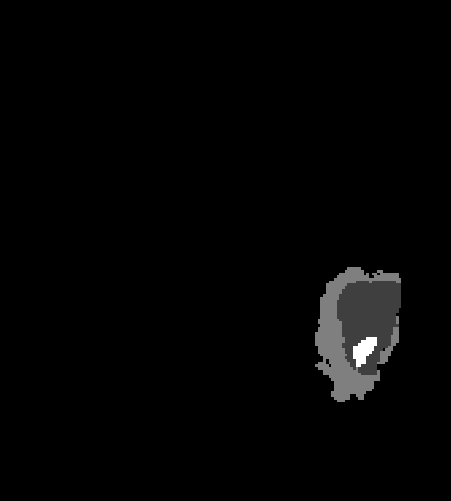}
\caption{}
\end{subfigure} \hfill
\centering
\begin{subfigure}{.21\textwidth}
\includegraphics[width=3cm,height=3cm]{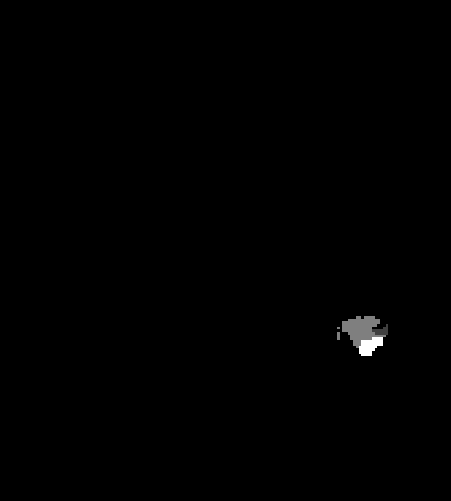}
\caption{}
\end{subfigure} \hfill
\centering
\begin{subfigure}{.21\textwidth}
\includegraphics[width=3cm,height=3cm]{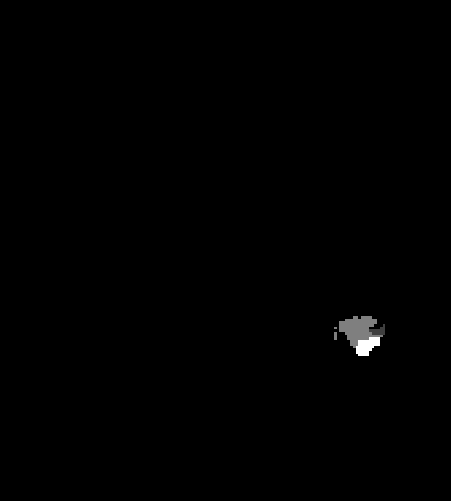}
\caption{}
\end{subfigure} 
\caption{Incorrect segmentation results of the network (a) FLAIR slice (b) Ground truth (c) Segmentation without post-processing (d) Segmentation after post-processing.}
\label{fig3}
\end{figure}

\begin{figure}[h]
	\begin{subfigure}{.4\textwidth}
		\centering
		\includegraphics[width=5.5cm,height=5.5cm]{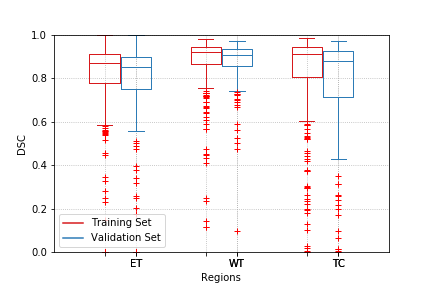}
		\caption{}
	\end{subfigure} \hfill
	\begin{subfigure}{.4\textwidth}
		\includegraphics[width=5.5cm,height=5.5cm]{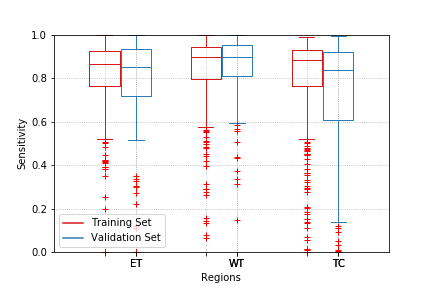}
		\centering
		\caption{}
	\end{subfigure} \vfill
	\begin{subfigure}{.4\textwidth}
		\includegraphics[width=5.5cm,height=5.5cm]{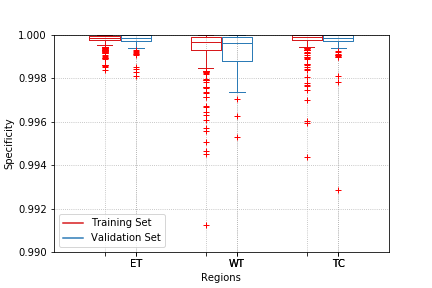}
		\centering
		\caption{}
	\end{subfigure} \hfill
	\begin{subfigure}{.4\textwidth}
		\includegraphics[width=5.5cm,height=5.5cm]{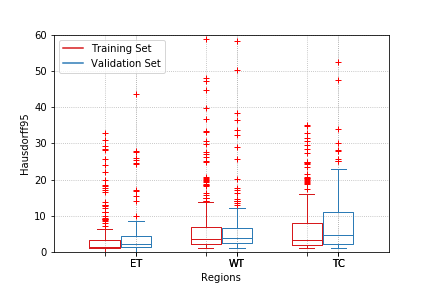}
		\centering
		\caption{}
	\end{subfigure}
	\caption{The box plot (a) DSC (b) Sensitivity (c)  Specificity (d)  Hausdorff95 }
	\label{fig4}
\end{figure}

The proposed model outperforms some of the ensemble approaches which is shown in Table \ref{tab9}. Fig. \ref{fig2} shows the successful segmentation of the tumor. The false positive segmentation voxels are removed in the post-processing. The network fails to segment the tumor for some HGG images and many LGG images. One such segmentation failure is shown in Fig. \ref{fig3}. The failure of the network is observed for: 1) small size of the entire tumor, 2) small size of necrosis, and 3) absence/small size of enhancing tumor. Fig. \ref{fig4} depicts the box plot of the evaluation metrics, where the red marked cases shows the segmentation failure. 

\subsection{OS Prediction}
The RFR trains on features extracted from the 213 ground truth images. In the trained RFR, features of network segmented images predict OS days. If the network fails to identify / segment necrosis from the image, then the feature extractor considers the absence of the necrosis and marks all the features except age as zero. OS accuracy on training, validation and test datasets for the images with GTR resection status is in Table \ref{tab7}. 

\begin{table}
\centering
\caption{OS accuracy for BraTS 2020 training, validation and test datasets.}
\label{tab7}
\begin{tabular}{p{2cm}p{2cm}p{2cm}p{2cm}p{2cm}p{2cm}}
\toprule
\textbf{Dataset} & \multicolumn{1}{c}{\textbf{Accuracy}} & \multicolumn{1}{c}{\textbf{MSE}} & \multicolumn{1}{c}{\textbf{MedianSE}} & \multicolumn{1}{c}{\textbf{StdSE}} &  \multicolumn{1}{c}{\textbf{SpearmanR}} \\ 
\midrule
\textbf{Training} & \multicolumn{1}{c}{0.568} &	\multicolumn{1}{c}{083165.963} &	\multicolumn{1}{c}{21481.525} &	\multicolumn{1}{c}{0181697.874} & \multicolumn{1}{c}{0.596} \\
\textbf{Validation} & \multicolumn{1}{c}{0.517} & \multicolumn{1}{c}{116083.477} & \multicolumn{1}{c}{43974.090} & \multicolumn{1}{c}{0168176.159} & \multicolumn{1}{c}{0.217} \\
\textbf{Test} & \multicolumn{1}{c}{0.477} & \multicolumn{1}{c}{382492.357} & \multicolumn{1}{c}{46612.810} & \multicolumn{1}{c}{1081670.063} & \multicolumn{1}{c}{0.333}\\
\bottomrule
\end{tabular}
\end{table}

The reduced performance of RFR on validation and test sets shows that it overfits the training dataset. Still, its performance is comparable with other approaches as shown in Table \ref{tab8}.

\begin{table}[!h]
\centering
\caption{Comparative analysis of OS accuracy on BraTS 2019 validation set.}
\label{tab8}
\begin{tabular}{c c c c}
	\toprule
	\textbf{Ref.} & \textbf{Approach} & \textbf{\# features} & \textbf{Accuracy (\%)} \\
	\midrule
	\cite{feng2019brain} & Linear Regression & 9 & 31.0\\
	\midrule
	\cite{pei2019multimodal} & Linear Regression & NA & 51.5 \\
	\midrule
	\cite{wang20193d} & Artifical Neural Network & 7 & 45.0 \\
	\midrule 
	\textbf{Proposed} & RFR & 18 & \textbf{51.7} \\
	\bottomrule	
\end{tabular}
\end{table}

According to the study \cite{sun2015integrative}, gender plays a vital role in response to tumor treatment. The females respond to the post-operative treatment better compared to males, which improves their life expectancy. The inclusion of the `gender’ feature into the existing feature list can significantly improve OS accuracy.

\section{Conclusion}
The proposal uses three-layer deep 3D U-net based encoder-decoder architecture for semantic segmentation. Each encoding and decoding layer module incorporates dense connections which allow the diversified feature learning and gradient propagation to the intial layers. The patch selection and loss function approaches improved the performance of the network. The pre-processing, post-processing and patch selection methods play vital role in the robust performance of the network. The network outperforms some of the state-of-the-art ensemble approaches. The network fails where the size of either the entire tumor or its subcomponent (necrosis and enhancing tumor) is comparatively small. The smaller subcomponent size is observed in LGG cases where the network fails significantly. The age, statistical, and necrosis shape features of the ground truth train RFR with five-fold cross-validation for OS prediction. Later, network segmentation for cases with GTR tests RFR for OS prediction. The RFR performs better than other state-of-the-art approaches which use linear regression and ANN.  

\section*{Acknowledgement}

The authors would like to thank NVIDIA Corporation for donating the Quadro K5200 and Quadro P5000 GPU used for this research, Dr. Krutarth Agravat (Medical Officer, Essar Ltd) for clearing our doubts related to medical concepts, Ujjawal Baid for his fruitful discussion in network selection. The authors acknowledge continuous support from Professor Sanjay Chaudhary, Professor N. Padmanabhan, and Professor Manjunath Joshi.

\bibliographystyle{splncs04}
\bibliography{bibtex_example}
\end{document}